\documentstyle[12pt]{article}
\begin{document}

\begin{center}
{\Large  Gravitational Instantons from Minimal Surfaces}\\[1cm]
{\Large A. N. Aliev}$^1$, {\Large M. Horta\c{c}su}$^{1,2},
${\Large J. Kalayc{\i}}$^{1,2}$ and {\Large Y. Nutku}$^1$ \\[1cm]
$^1$ Feza G\"ursey Institute, P. O. Box 6 Cengelkoy, 81220 Istanbul, Turkey
\\$^2$Istanbul Technical University, Department of Physics, 80626 Maslak
Istanbul, Turkey\\
\end{center}

Physical properties of gravitational instantons which are derivable from
minimal surfaces in $3$-dimensional Euclidean space are examined using the
Newman-Penrose formalism for Euclidean signature. The gravitational
instanton that corresponds to the helicoid minimal surface is investigated
in detail. This is a metric of Bianchi
Type $VII_0$, or E(2) which admits a hidden symmetry due to the existence of
a quadratic Killing tensor. It leads to a complete separation of variables
in the Hamilton-Jacobi equation for geodesics, as well as in Laplace's
equation for a massless scalar field.
The scalar Green function can be obtained in closed form which enables us
to calculate the vacuum fluctuations of a massless scalar field in the
background of this instanton.

\section{Introduction}

The interest in instantons originates from the discovery of finite-action
solutions of the classical Yang-Mills equations, instantons
\cite{bpst}-\cite{jnr},
which are localized in imaginary time. They provide the dominant
contribution to the path-integral in the quantization of the Yang-Mills
fields. The expectation that their gravitational counterparts should play a
similar role in the path-integral approach to quantum gravity has been
a constant stimulus for research on gravitational instantons.
Recently gravitational instantons which are described by hyper-K\"ahler
metrics were studied extensively in the framework of supergravity
and $M$-theory as well as Seiberg-Witten theory \cite{ket2}-\cite{cihan}.

Gravitational instantons are given by regular complete metrics with
Euclidean signature and self-dual curvature which implies that they satisfy
the vacuum Einstein field equations \cite{h}-\cite{ahs}. The simplest
examples of gravitational instantons are obtained from the
Schwarzschild-Kerr and Taub-NUT solutions by analytically continuing them to
the Euclidean sector \cite{h},\cite{gh}. Through this construction we obtain
nonsingular positive-definite metrics which are asymptotically flat at
spatial infinity and periodic in imaginary time. These properties play a
central role in the Euclidean path-integral approach to the quantum
mechanics of black holes \cite{gh1}. The principal class of
physically interesting gravitational instantons consist of
Gibbons-Hawking multi Taub-NUT metrics \cite{gh} which are
asymp\-toticaly locally Euclidean metrics with self-dual curvature.
The Eguchi-Hanson instanton \cite{eh},\cite{egh} is a distinguished
member of this class.

In the Euclidean path-integral approach to quantum gravity one is interested
in the evaluation of the functional integral over all stationary phase
metrics with appropriate boundary conditions. Gravitational instantons
should provide the dominant contribution to the path integral and mediate
the quantum tunnelling between two homotopically distinct vacua. Therefore
the search for new instanton solutions of the Einstein field equations is of
great interest physically, as well as mathematically. Recently it was
observed \cite{min} that minimal surfaces in Euclidean space can be used in
the construction of instanton solutions, even as in the case of Yang-Mills
instantons \cite{comtet}. For every minimal surface
in 3-dimensional Euclidean space there exists a gravitational
instanton which is an exact solution of the Einstein field equations with
Euclidean signature and self-dual curvature. If the surface is defined by
the Monge ansatz $\phi=\phi(x,t)$, then the metric establishing this
correspondence is given by 
\begin{eqnarray}
\label{ansatz1}
ds^2 & = & \frac{1}{\sqrt{1+\phi_{t}^{\;2}+\phi_{x}^{\;2} }} \left[
( 1 + \phi_{t}^{\;2} ) ( d t^2 + d y^2 ) 
\right. \\  & & \left. \nonumber
+ ( 1+\phi_{x}^{\;2} ) ( d x^2 + d z^2 )
+ 2 \phi_{t}\,\phi_{x} ( d t \, d x + d y \, d z )  \right]
\end{eqnarray}
whereby the Einstein field equations reduce to the classical equation 
\begin{equation}
\label{minimal}(1+\phi _x^{\;2})\,\phi _{tt}-2\,\phi _t\,\phi _x\,\phi
_{tx}+(1+\phi _t^{\;2})\,\phi _{xx}=0
\end{equation}
governing minimal surfaces in $R^3$. These are K\"ahler metrics obtained
from J\"orgens' correspondence \cite{jorgens} between the equation for
minimal surfaces and the real elliptic Monge-Amp\`ere equation in $2$%
-dimensions. Gravitational instantons that follow from this construction
will admit at least two commuting Killing vectors $\partial _y,\partial _z$
which implies that they may be complete non-compact Ricci-flat K\"ahler
metrics which have been considered in the context of stringy cosmic strings
\cite{gsvy}, \cite{gorr}. The general gravitational instanton metric that
results from Weierstrass' general local solution \cite{weier} for minimal
surfaces has been constructed in \cite{akn} using the correspondence
(\ref{ansatz1}) between minimal surfaces and gravitational instantons.

In this paper we shall use the Newman-Penrose formalism for Euclidean
signature developed in \cite{previous}, which will henceforth be
referred to as {\bf I}, to investigate the physical properties
of gravitational instantons which are derivable from minimal surfaces.
Among all such solutions, the gravitational instanton that corresponds
to the helicoid minimal surface is of particular interest, however, it
is incomplete and has a curvature singularity. The physical interpretation
of this metric as an instanton must therefore await an analysis of its
global structure. In this connection it is worth noting that singular
hyper-K\"ahler instanton metrics sometimes admit an $M$-theory
resolution \cite{ket2} and are therefore of interest in supergravity.
It is possible that a supergravity extension of the metric corresponding
to helicoid may also be of interest and therefore we shall
investigate its properties in some detail. We shall show that the self-dual
metric derived from helicoid has a number of remarkable properties. It admits
a 3-parameter group of motions $E(2)$, namely rotation and boosts on the
Euclidean plane, which is also known as $G_3$, and the metric is of Bianchi
Type $VII_0$. Furthermore, it admits a hidden symmetry
which stems from the existence of a quadratic Killing
tensor. Next we shall show that the Hamilton-Jacobi equation for geodesics,
as well as Laplace's equation for a massless scalar field are
separable in the background of the metric generated by the helicoid.
We construct the scalar Green function in closed form and calculate the
vacuum fluctuations of a massless scalar field in the background of this
instanton. Using the point-splitting procedure we obtain the renormalized
expression for vacuum expectation value of the stress-energy tensor.

\section{General minimal surface solution}

The instanton metric obtained from Weierstrass' general local solution for
minimal surfaces is given by \cite{akn} 
\begin{equation}
\label{gms}ds^2=(\,1-|\,g\,|^4\,)\,|\,f\,|^2\,d\zeta \,d\bar \zeta +\frac
1{1-|\,g\,|^4}\,|\;d\sigma -\bar g^2d\bar \sigma \;|^{\;2}
\end{equation}
where $\sigma =y+iz$ is a complex coordinate and $f,g$ are arbitrary
holomorphic functions of $\zeta $ which replaces $t,x$ as a new complex
coordinate through the Weierstrass formulae \cite{weier}, \cite{akn}.
The metric (\ref{gms}) contains two arbitrary holomorphic functions but
only one of them, namely $g$, is geometrically significant.
Nevertheless, for some purposes it may be useful to keep $f$.

In order to investigate the geometrical properties of the
metric (\ref{gms}) we shall use the Newman-Penrose formalism for Euclidean
signature and refer to the results of {\bf I}. For this purpose we first
need to specify a tetrad which will be given as an isotropic complex
dyad defined by the vectors $l$, $m$ together with
their complex conjugates subject to the normalization conditions 
\begin{equation}
l_{\mu} {\bar l}^{\mu}=1 \;\;\;\;\;\;\;\;m_{\mu} {\bar m}^{\mu}=1
\label{norm}
\end{equation}
with all others vanishing. The power of the Newman-Penrose formalism becomes
evident if the legs of this complex dyad are chosen along an isotropic
geodesic congruence. Using eqs.({\bf I}.19) and ({\bf I}.20) we obtain
\begin{eqnarray}
l_{\mu; \,\nu} l^{\nu} & = & (\epsilon -\bar{\gamma}) \,l_{\mu} +
\kappa \, {\bar m}_{\mu} - \bar \nu \, m_{\mu}
\label{nullgeod1}           \\[2mm]
m_{\mu; \,\nu} m^{\nu} & = & (\bar \alpha -\beta) \,m_{\mu} +
\mu \, l_{\mu} - \sigma \, \bar{l}_{\mu}
\label{nullgeod2}
\end{eqnarray}
where semicolon denotes covariant differentiation. From these equations it
follows that if the spin coefficients $\kappa=\nu=0$, or $\mu=\sigma=0$,
then the vector fields with components $l^{\mu}$, or $m^{\mu}$ determine the
corresponding isotropic geodesic congruence up to affine reparametrization.

For the instanton metric (\ref{gms}) the obvious co-frame is given 
\begin{eqnarray}
\label{nulltetrad}
l & = & \frac{1}{\sqrt{2}} ( 1 - |\, g\, |^4  )^{1/2} \, |\, f\, | \,
d \zeta,  \\ [2mm]
m & = & \frac{1}{\sqrt{2}} (1 - | \, g\, |^4 )^{-1/2} \, \left( \,
  d \sigma - \bar{g}^2 \, d \bar{\sigma}   \, \right) \nonumber
\label{coframe}
\end{eqnarray}
and its inverse will provide a convenient choice of the complex dyad.
In this case there exists an isotropic geodesic congruence formed by only one
basis vector, namely $l^\mu $. To make this explicit we take the exterior
derivative of the basis 1-forms (\ref{coframe}) and compare the result with
eqs.({\bf I}.21) to obtain the spin coefficients for the metric (\ref{gms}).
Thus we have
\begin{eqnarray}
\kappa & = & \nu = \alpha = \beta = \pi= \tau = \mu = \lambda = \rho = 0 ,
\nonumber \\ [2mm]
\epsilon & = & \frac{1}{2\sqrt{2}\,|\, f\, |^3}
 (1 - | \, g\, |^4 )^{-3/2} \,
 \left[ (1 - | \, g\, |^4 ) \, f \, \bar{f}' - 4 |\, f\, |^2 \, |\, g\, |^2
\,  g \, \bar{g}' \, \right],    \nonumber   \\ [2mm]
\gamma & = & - \frac{1}{2\sqrt{2}\, |\, f\, |^3} (1 - | \, g\, |^4 )^{-1/2} \,
f \,  \bar{f}' , \label{spincoeff}  \\  [2mm]
\sigma & = & - 2 \sqrt{2} \, |\, f\, |^{-1} (1 - | \, g\, |^4 )^{-3/2} \,
  \bar{g} \, \bar{g}' ,  \nonumber
\end{eqnarray}
where prime denotes derivative with respect to the argument. We see that the
spin coefficients $\kappa $ and $\nu $ vanish so that the vector field
$ {\bf l} $ forms an isotropic geodesic congruence. We note that
for the choice $f=1$ the spin coefficient $\gamma $ also vanishes
resulting in an anti-self-dual gauge,
{\it cf} eqs.({\bf I}.45). Next, using the spin coefficients
(\ref{spincoeff}) in the Ricci identities ({\bf I}.93,95) we
find that the set of Weyl scalars labelled with tilde vanish $\tilde
\Psi _i=0,\; i=0,1,..,4 \; $ reflecting the anti-self duality of
the curvature tensor. For the remaining set of Weyl scalars we find
\begin{eqnarray}
\Psi_{1} & = & - \bar{\Psi}_{3} = 0 , \nonumber \\ [2mm]
\Psi_{2} & = & \frac{8}{|\, f\, |^2 \,(1 - | \, g\, |^4 )^{3}} \,
 |\,g \, \bar{g}'\,|^2  \label{weylsgen} \\ [2mm]
\Psi_{0} & = & \bar{\Psi}_{4} =
- \frac{4}{|\, f\, |^2 \,(1 - | \, g\, |^4 )^{2}} \,
 \left[\bar{g} \, \bar{g}''  -  \frac{f \,  \bar{f}'}{|\, f\, |^{2}}\,\,
 \bar{g} \, \bar{g}' +\frac{1 + 5 \, | \, g\, |^4}{1 - | \, g\, |^4}\,\,
 (\bar{g}')^2  \right]  \nonumber
\end{eqnarray}
which shows that, {\it cf} eqs.({\bf I}.121), the instanton metric
(\ref{gms}) obtained from Weierstrass' formulae for minimal
surfaces is of Petrov Type $\,I\,$. Finally, we note that
\begin{equation}
\label{sing}|\,g\,|^4=1
\end{equation}
defines the locus of curvature singularities.

\section{Helicoid}

Among the various particular realizations of the general metric (\ref{gms}),
the instanton metric that corresponds to the helicoid minimal surface
appears to be the one with greatest interest. The graph of
the helicoid is given by
\begin{equation}
\label{cathelsurf}\phi =a\;\tan ^{-1}\left( \frac xt\right) 
\end{equation}
and by introducing new coordinates
\begin{equation}
\label{rt}x=r\cos \theta,\;\;\;\;t=r\sin \theta,
\end{equation}
the corresponding instanton metric becomes \cite{min} 
\begin{eqnarray}
d s^2  & = &  \frac{1}{\sqrt{ 1 + \frac{\textstyle{a^2}}{\textstyle{r^2}}}}
 \left[   d r^2  +  (r^2 + a^2)  \, d \theta^2
   +  \left( 1 + \frac{a^2}{r^2}  \sin^2  \theta \right) d y^2
\right. \nonumber \\  & & \left.
\mbox{}  -  \frac{a^2}{r^2}  \sin 2 \theta \,  d y \, d z
 + \left( 1 + \frac{a^2}{r^2} \cos^2  \theta \right)  d z^2    \right]
\label{ch}
\end{eqnarray}
where the coordinates $ y $  and $ z $ along the Killing directions
will be taken to be periodic, coordinates on a $2$-torus,
as in the discussion of stringy cosmic strings \cite{gsvy}.
There are two asymptotic regions
$ r \rightarrow \pm \infty$ which must be identified.
The singularity in the metric (\ref{ch}) at $r=0$ is a curvature
singularity because we shall show that curvature scalars are singular there.
In this paper we shall not discuss the global properties of the metric
(\ref{ch}) but instead point out some of its remarkable properties such as
a hidden symmetry in addition to Bianchi $VII_0$ symmetry. This is
described by a Killing tensor and leads to a complete
separation of variables in the Hamilton-Jacobi equation for geodesics,
as well as in Laplace's equation for a massless scalar field.

The coordinates $r, \theta$ are inherited from the standard description of
the helicoid and we can immediately recognize the first fundamental
form of the helicoid $ I = d r^2 + (r^2 + a^2) \, d \theta^2 $ in
the metric (\ref{ch}). The helicoid is a ruled surface. There are further
helicoids here, namely the constant $r=r_0, z=z_0$ sections of the
metric (\ref{ch})
\begin{eqnarray}
d l^2   =  (r_{0}^2 + a^2)  \, d \theta^2
   +  \left( 1 + \frac{a^2}{r_{0}^2}  \sin^2  \theta \right) d y^2
\label{rzconst}
\end{eqnarray}
are helicoids in the torus $S^2 \times S^1$ as shown in Fig. 1  and
similarly, the $2$-sections defined by constant $r, y$ are also helicoid
surfaces in another $S^2 \times S^1$. These helicoids are ruled by
a geodesic of $S^2$.

   An alternative coordinate system for the metric (\ref{ch})
obtained by letting
\begin{equation}
\label{defchi}   r = a \, \sinh \chi
\end{equation}
results in the following form of the metric
\begin{eqnarray}
d s^2  & = &  \frac{a^2}{2} \sinh 2\chi \,
 ( d \chi^2  +  d \theta^2 )
\,+ \frac{2}{\sinh 2\chi} \,
\left[ (\sinh^2 \chi +\sin^2 \theta)\,dy^2
\nonumber \right. \\ [2mm]  & & \left. - \sin2 \theta \, dy dz
+\, (\sinh^2 \chi +\cos^2 \theta )\,dz^2 \right]
\label{helmet1}
\end{eqnarray}
which can be obtained directly from the
general minimal surface solution (\ref{gms}) through the choice
of holomorphic functions
\begin{equation}
\label{holfunc}f=a\,e^\zeta,\;\;\;\;\;\;g=e^{-\zeta }
\end{equation}
where $\zeta =\chi +i\theta $.

   The metric (\ref{helmet1}) is of Bianchi Type $VII_0$. It is a homogeneous
anisotropic manifold that admits a $3$-parameter group of motions
which is the same as $E(2)$, the group of motions on the Euclidean plane.
The metric (\ref{helmet1}) can therefore be written in the compact form
\begin{equation}
\label{metric}
d s^2 = \frac{a^2}{2} \sinh 2 \chi \, \left[ d \chi^2 + (\sigma^3)^{2}
   \right] +  \tanh \chi \, (\sigma^1)^{2}  +  \coth \chi \, (\sigma^2)^{2}
\end{equation}
using the left-invariant $1$-forms
\begin{eqnarray}
\sigma^1 & = &  \cos \theta \,d y + \sin \theta \, d z, \nonumber \\
\sigma^2 & = &  - \sin \theta \,d y + \cos \theta \, d z, \label{left6} \\
\sigma^3 & = & d \theta, \nonumber
\end{eqnarray}
of Bianchi Type $VII_0$ \cite{mac}. They satisfy the
Maurer-Cartan equations of structure
\begin{equation}
\label{mc}
d \sigma^i = \frac{1}{2} c_{j\;\;k}^{\;\;i} \; \sigma^j \wedge \sigma^k
\end{equation}
where
\begin{equation}
\label{strconst}
c_{3\;\;2}^{\;\;1} = - c_{2\;\;3}^{\;\;1} = c_{1\;\;3}^{\;\;2} = -
c_{3\;\;1}^{\;\;2} = 1
\end{equation}
are the only non-vanishing structure constants.
A different representation of the metric (\ref{metric}) is obtained if
in place of the left-invariant $1$-forms $\sigma^i$ we use the
right-invariant $1$-forms of Bianchi Type $VII_0$
\begin{eqnarray}
r^1 & = & d y - z \, d \theta, \nonumber \\
r^2 & = & d z + y \, d \theta, \label{lefte2} \\
r^3 & = & d \theta \nonumber
\end{eqnarray}
and it can be directly verified that the metric
\begin{equation}
\label{metricr}
d s^2 = \frac{a^2}{2} \sinh 2 \chi \, \left[ d \chi^2 + ( r^3)^{2}
   \right] +  \tanh \chi \, ( r^1)^{2}  +  \coth \chi \, ( r^2)^{2}
\end{equation}
is Ricci-flat and anti-self-dual as well.
Even though the explicit expression for the metric changes
drastically under this exchange of left and right-invariant $1$-forms,
there will be no change in the principal results we shall
present below.

   In order to investigate the geometrical properties of the metric
(\ref{metric}) we shall use the Newman-Penrose formalism for
Euclidean signature. For this metric the natural complex
isotropic dyad is given by the co-frame
\begin{eqnarray}
l  & = &  \frac{a}{2} \left(\sinh 2\chi \right)^{1/2}
 \left(\,d \chi   +  i \,   \sigma^3 \,\right)
 \nonumber \\ [2mm] \label{tetrad1}
 m & = &  \frac{1}{\sqrt{2}} \left[ \left( \tanh \chi \right)^{1/2} \sigma^1
 +  i \left( \coth \chi \right)^{1/2}  \sigma^2 \right] \nonumber
\end{eqnarray}
and this choice results in an anti-self-dual gauge which can be
verified by calculating the spin coefficients through eqs.({\bf I}.21).
We find
\begin{equation}
\epsilon  =  \frac{1}{a}\, \cosh 2 \chi \, (\sinh 2\chi)^{-3/2},
\;\;\;\;\;\;\;
\sigma  =   \frac{2}{a}\, (\sinh 2 \chi)^{-3/2}
\label{spincoeff1}
\end{equation}
while all other spin coefficients vanish. Now from eq.(\ref{nullgeod1}) it
is readily seen that the vector field $\,{\bf l} \,$
determines an isotropic geodesic congruence up to affine reparametrization.
Using the spin coefficients in eqs.({\bf I}.36) we find that the
connection 1-forms are given by
\begin{eqnarray}
\Gamma_{0}^{\;\;0}=-\Gamma_{1}^{\;\;1} & = &
 \frac{1}{a}\, \cosh 2 \chi \,(\sinh 2\chi)^{-3/2}\,\,(\bar{l} - l )
\nonumber \\[2mm]
\Gamma_{0}^{\;\;1}=-\bar{\Gamma}_{1}^{\;\;0}
& = &- \frac{2}{a}\, (\sinh 2\chi)^{-3/2} \,\, \bar{m}
\label{potform}
\end{eqnarray}
and in particular all $\tilde{\Gamma}_{x'}^{\;\;y'}$ vanish so that this
is an anti-self-dual gauge. The curvature tetrad scalars are obtained
through the Ricci identities by substituting these spin coefficients in
eqs.({\bf I}.93, 95) and we find that the non-vanishing Weyl scalars
are given by
\begin{eqnarray}
\Psi_{2}  & = &  \frac{4}{a^2} \,(\sinh 2\chi)^{-3},  \label{weyls1} \\
\Psi_{0} & = & \bar{\Psi}_4 =
  - \frac{12}{a^2}\, \cosh 2\chi\, (\sinh 2\chi)^{-3},  \nonumber
\end{eqnarray}
thus the curvature is anti-self-dual, the metric is algebraically general,
or Petrov Type $I$ \cite{previous} and there is a curvature singularity
$ r=0 $, or $ \chi=0$.
We conclude that the instanton metric that corresponds to
the catenoid minimal surface, where $ a^2 \rightarrow - a^2$ in
eq.(\ref{ch}), is Bianchi Type $VI_0$. Both of these metrics can be
obtained from the Belinski-Gibbons-Page-Pope Bianchi Type $IX$ and
$VIII$ solutions \cite{bgpp} by group contraction \cite{iw} which
requires that the modulus of the elliptic functions must be set
equal to unity.

\section{Symmetries}

Well-known gravitational instanton metrics admit hidden symmetries \cite{gru}
and we shall now show that the instanton metric derived from
helicoid minimal surface provides another such example. First of all
the metric (\ref{metric}) admits a three parameter group of motions.
The Killing vectors are given by
\begin{equation}
\xi_1= \frac{\partial}{\partial y}, \;\;\;\;\;
\xi_2 = \frac{\partial}{\partial z}, \;\;\;\;\;
\xi_3= \frac{\partial}{\partial \theta}
- z\, \frac{\partial}{\partial y}
+ y\, \frac{\partial}{\partial z}
\label{3kill}
\end{equation}
and they satisfy the Lie algebra
\begin{equation}
\left[\xi_1, \xi_2 \right] = 0 , \;\;\;\;\;
\left[\xi_1, \xi_3 \right] =  \xi_2 , \;\;\;\;\;
\left[\xi_3, \xi_2 \right] =  \xi_1
\label{algebra}
\end{equation}
of Bianchi Type $VII_0$ with the structure constants
(\ref{strconst}). The Killing vectors (\ref{3kill}) are the right-invariant
vector fields satisfying
\begin{equation}
i_{\xi_k} \, r^j = \delta^{j}_{k}
\label{comp}
\end{equation}
upon contraction with the right-invariant $1$-forms (\ref{lefte2}).

  Hidden symmetries play an important role in the study of first integrals
of motion. Starting from the Lagrangian for geodesics
$$ L=\frac 12\left( \frac{ds}{d\lambda }\right) ^2 $$ where
$\lambda $ is an affine parameter, we have the conserved canonical momenta
\begin{eqnarray}
p_{y}= \frac{2}{\sinh 2 \chi}\,
\left[  (\sinh^2 \chi +\sin^2 \theta)\,\dot{y}
\,- \,\sin \theta \,\cos \theta \,\dot{z} \right]
\\[2mm] \nonumber
p_{z}= \frac{2}{\sinh 2 \chi}\,
\left[  (\sinh^2 \chi +\cos^2 \theta)\,\dot{z}
\,- \,\sin \theta \,\cos \theta \,\dot{y} \right]
\label{momenta}
\end{eqnarray}
conjugate to $\,y\,$ and $\,z\,$ respectively,
where dot denotes differentiation with respect to $\lambda $. Since the
metric (\ref{metric}) admits two evident Killing vectors
$\partial _y $ and $\partial _z$, these two first integrals of motion
are manifest. Another immediate constant of motion is the square of
the 4-momentum
\begin{equation}
g^{\mu \nu }\,p_\mu \,p_\nu = \mu^2 ,
\label{sq4mom}
\end{equation}
however, these three integrals of motion are not sufficient for a complete
integration of the geodesic equations. Thus, following Walker and Penrose
\cite{wp} we shall consider possible quadratic integrals of motion
\begin{equation}
K_{i}= K_{i}^{\mu \nu }\,p_\mu \,p_\nu
\label{4thint}
\end{equation}
where $ K_{i}^{\mu \nu } $ are symmetric Killing tensors which
satisfy the Walker-Penrose equations
\begin{equation}
\label{killing}K_{i}^{(\mu \nu ;\,\lambda )}=0
\end{equation}
and round parentheses denote symmetrization.
It is evident that the existence of a non-trivial Killing tensor in the
instanton metric (\ref{metric}) should give rise to a fourth integral
of motion. In order to integrate eqs.(\ref{killing}) we shall use
the Newman-Penrose formalism for Euclidean signature. We begin
with the resolution of the Killing tensor along the legs of
the complex isotropic dyad (\ref{tetrad1})
\begin{eqnarray}
 K^{\mu \nu} & = & \chi_{11}\, \bar l^{\mu} \bar l^{\nu} + \chi_{33} \,
 \bar m^{\mu} \bar m^{\nu} + \, \chi_{12} \, \bar l^{\mu} l^{\nu} + \chi_{34}
\, \bar m^{\mu} m^{\nu}
 \nonumber \\ [2mm] & &
+ 2\,\chi_{13}\, \bar l^{\,(\mu} \bar m^{\nu)}
 + 2 \chi_{14} \,\bar l^{\,(\mu} m^{\nu)}  + \, cc
\label{killtenexp}
\end{eqnarray}
where $ cc $ denotes the complex conjugate of the foregoing terms
and the Killing tetrad scalars will be defined as
\begin{eqnarray}
\chi_{11}  & = & \bar \chi_{22}  =  K_{\mu \nu} l^{\mu}l^{\nu}
\nonumber \\
\chi_{33} & = & \bar \chi_{44}   =  K_{\mu \nu} m^{\mu} m^{\nu}
\nonumber \\
\chi_{13}  & = & \bar \chi_{24}  =  K_{\mu \nu} l^{\mu} m^{\nu}
\label{killscal} \\
\chi_{14} & = & \bar \chi_{23}   =  K_{\mu \nu} l^{\mu} \bar m^{\nu}
 \nonumber \\
\chi_{12}  & = & \bar \chi_{21}  =  K_{\mu \nu} l^{\mu} \bar l^{\nu}
\nonumber \\
\chi_{34}  & = & \bar \chi_{43}  =  K_{\mu \nu} m^{\mu} \bar m^{\nu} .
\nonumber
\end{eqnarray}
From eqs.(\ref{killing}) immediately it follows that
the dyad scalars $ \chi_{13} $ and $ \chi_{14} $ must vanish
while the remaining satisfy a set of coupled equations
\begin{eqnarray}
D\chi _{11}-2\,\epsilon \,\chi _{11}&=&0, \nonumber \\
(\, \bar D + 2\,\epsilon \,) \,\chi _{11}+ 2\,D\chi _{12}&=&0,  \nonumber \\
(\, D + 2\,\epsilon \,) \,\chi _{44}+ 2\,\sigma \,\chi _{11} &=&0,
\label{npkiltens} \\
(\, D - 2\,\epsilon \,) \,\chi _{33} + 2\,\sigma (\chi _{12} - \chi_{34})
&=&0, \nonumber \\
D\chi _{34}-\sigma \,\chi _{44}&=&0  \nonumber
\end{eqnarray}
which admit the general solution
\begin{eqnarray}
\chi_{11}  & = & \chi_{22}  =
A \,\sinh\chi\, \cosh\chi ,\;\;\;\;
\chi_{12}   =  -\chi_{11} + B,  \nonumber   \\ [2mm]
\chi_{33} & = & \chi_{44}
  = a^2\,\tanh \chi \,\left( -A + \frac{C}{a^4\,\sinh^2 \chi}\right),
\label{killscal1} \\ [2mm]
\chi_{34}  & = & - \frac{a^2}{\sinh\chi\, \cosh\chi}
\,\left( A\,\sinh^2 \chi  + \frac{C}{a^4}\,\cosh 2\chi \right) \, +\, B
\nonumber
\end{eqnarray}
with arbitrary constants $A,\,B $ and $ C $. Substituting these
expressions into eq.(\ref{killtenexp}) we find that the additive
constants of integration  $\,B \,$ and  $ C $ do not
give rise to new Killing tensors. They lead to constants of motion
derived from the metric tensor and the symmetric tensor product
of the two commuting Killing vectors respectively. Therefore
without loss of generality we can discard them and we
are left with the independent integral of motion
\begin{equation}
K   =  -\,{p_\theta}^2 \,-
\,a^2\,\left( \cos \theta \, p_y + \sin \theta \, p_z \right)^2
\label{integral}
\end{equation}
which is not expressible in terms of
symmetric bilinear combinations of the Killing vectors (\ref{3kill}).
Thus the instanton metric that corresponds to the helicoid minimal
surface admits the quadratic Killing tensor (\ref{integral})
which provides a fourth integral of motion for geodesics.

The existence of this Killing tensor leads to a separation of variables in
the Hamilton-Jacobi equation for geodesics. The separability ansatz
\begin{equation}
\label{action}S=p_y\,y\,+\,p_z\,z\,+\,S_{\chi}(\chi)\,+\,S_\theta (\theta )
\end{equation}
leads to decoupled ordinary differential equations in the Hamilton-Jacobi
equation
\begin{eqnarray}
 \left(\frac{d S_{\chi}}{d \chi} \right)^2
+a^2\,\left( p_{y}^2 + p_{z}^2 \right) \,\sinh^2 \chi -
 \frac{ \mu^2\,a^2}{2}\,\sinh 2\chi = K
\nonumber  \\ [2mm]
\left(\frac{d S_{\theta}}{d \theta} \right)^2 +
\, a^2\,\left( \cos \theta\, p_{y}
 \,+\, \sin \theta p_{z}\right)^2 = - K
\label{hj}
\end{eqnarray}
where the separation constant $ K $ coincides with (\ref{integral}).
As it was noted in \cite{gru} some well-known instanton metrics with
two commuting Killing vectors which, however, are not of Petrov type
$ D $ may admit a Killing tensor. We have seen that the metric
(\ref{metric}) which is  an anti-self-dual Petrov type $ I $ instanton
provides another example of such metrics.

\section{Scalar Green's function}

The Green function for a massless scalar field obeys the equation
\begin{equation}
\label{green }\Delta G ( x,x') = - \delta^4 (x,x^{\prime})
\end{equation}
where $\Delta = \nabla_{\mu} g^{\mu\nu} \nabla_{\nu}$ is the Laplacian
with $\nabla$ denoting the covariant derivative. For the metric
(\ref{metric}) it has the following explicit form
\begin{eqnarray}
\Delta & = & {\partial_{\chi\chi}} + {\partial_{\theta\theta}}
+ a^2\,\sinh^2 \chi\, \left({\partial_{yy}} + {\partial_{zz}} \right)
\nonumber  \\[2mm] & &
 + \,\,a^2 \left( \cos \theta \,\partial_y  + \sin \theta \,
 \partial_z \right)^2
\label{laps}
\end{eqnarray}
and we shall now show that the existence of a Killing tensor leads to the
solution of Laplace's equation by separation of variables which enables us
to construct the scalar Green function in closed form. Following Carter
\cite{car1} we begin with the second order operator which is constructed
from the Killing tensor (\ref{integral})
\begin{equation}
 \hat K= \nabla_{\mu} K^{\mu\nu} \nabla_{\nu}= - \, {\partial_{\theta\theta}}
 - \,a^2 \left( \cos \theta \,\partial_y  + \sin \theta \,
 \partial_z \right)^2
\label{Koper}
\end{equation}
which commutes with the Laplacian (\ref{laps}) and the vector fields
$\partial_y $ and $\partial_z $. Thus the three commuting operators
\begin{eqnarray}
\partial_y \Phi & = & k_y\, \Phi         \nonumber \\
\partial_z \Phi & = & k_z\, \Phi  \label{eigenf} \\
\hat K \Phi & = &\lambda \,\Phi \nonumber
\end{eqnarray}
with eigenvalues $ k_y,\, k_z, \, \lambda $ have common eigenfunctions.
Hence eigenfunctions of the Laplacian  (\ref{laps})
admit separation of variables of the form
\begin{equation}
\Phi =   R(\chi) \, S(\theta) \, e^{i( k_y  y  \, + \, k_z  z )}
\label{generic}
\end{equation}
where the angular functions satisfy the last one of eqs.(\ref{eigenf}).
This can be reduced to a pair of Mathieu equations by passing to polar
coordinates in the constants of separation
$ k_y = k \cos\phi$ and $ k_z = k \sin\phi$. Then we have
\begin{equation}
\frac{d^2 S }{d \Theta^{2}}  +  \left(\, \lambda -
 k^2 \,a^2 \cos^2 \Theta \,\right) S = 0
\label{mathi1}
\end{equation}
where $ \Theta = \theta - \phi $.  We are interested only in periodic
solutions of this equation with period $\,2\pi \,$. These solutions
exist only for discrete values of the separation constant $ \,\lambda\, $
and they are given by even and odd periodic Mathieu functions
$ Se_{n}( ka,\, \cos\Theta ) $ and $ So_{n}( ka,\, \cos\Theta ) $
respectively \cite{mf}. When the parameter $  ka  $ tends to zero, these
solutions reduce to the trigonometric functions
$$ Se_{n}( ka,\, \cos\Theta ) \rightarrow \;\cos(n \Theta) ,\;\;\;\;\;
So_{n}( ka,\, \cos\Theta ) \rightarrow \;\sin(n \Theta) $$
while the separation constant
$$ \lambda e_{n}   \rightarrow \; \lambda o_{n} \rightarrow \; n^2 $$
goes over into the square of an integer.

Similarly, the equation for radial modes can also be transformed into
Mathieu's equation
\begin{equation}
\frac{d^2 R}{d \tilde \chi ^2}  + ( k^2\, a^2
\cosh^2 \tilde \chi -  \lambda )\, R  = 0
\label{mathi2}
\end{equation}
where
$$  \tilde{\chi} = \chi+i\frac{\pi}{2} $$
is a new complex coordinate. The solutions of this equation which
satisfy the regularity conditions at the origin and at infinity
are expressed in terms of Bessel-like Mathieu functions
$Je_{n} ( ka, \cosh \tilde \chi),$
$Jo_{n} ( ka, \cosh \tilde \chi)$
and Hankel-like Mathieu functions $He_{n} ( ka, \cosh \tilde \chi),$
$Ho_{n} ( ka, \cosh \tilde \chi)$ respectively.
With these solutions of eqs.(\ref{mathi1})-(\ref{mathi2}) we use the
general procedure \cite{mf} to obtain the Green function
\begin{equation}
 G ( x,x') = \frac{1}{(2\pi)^2} \int_0^{\infty} \int_0^{2\pi} k \,
dk \, \, d \phi \; e^{i k Y \cos \phi}\,
g_h (\tilde \chi,\tilde \chi^{\prime},\Theta,\Theta^{\prime})
\label{green1}
\end{equation}
with the factorized function
\begin{eqnarray}
g_h (\tilde \chi,\tilde \chi^{\prime},\Theta,\Theta^{\prime}) & = &
i \pi H_0^{(1)}(k \tilde{Z})=   4 \pi i \left\{
 \sum _{n=0}^{\infty} \left(\frac{Se_{n}
(h,\cos \Theta')}{Me_{n}(h)} \right) Se_{n}( h,\cos \Theta ) \right.
\nonumber \\  & & \hspace{5mm}         \times
\left[ \theta(\tilde{\chi}-\tilde{\chi}')
Je_{n}(h,\cosh \tilde{\chi}') \,He_{n}(h,\cosh \tilde{\chi})
\right. \nonumber  \\  & & \left. \hspace{1cm}
+\theta(\tilde{\chi}'-\tilde{\chi})
Je_{n}(h,\cosh \tilde{\chi})\, He_{n}(h,\cosh \tilde{\chi}') \right]
\nonumber   \\ & & + \sum_{n=0}^\infty \left(\frac{So_{n}
(h,\cos \Theta')}{Mo_{n}(h)} \right)
So_{n} (h,\cos \Theta)
\nonumber    \\  & & \hspace{5mm}    \times
\left[ \theta(\tilde{\chi}-\tilde{\chi}')
Jo_{n} (h,\cosh \tilde{\chi}')\, Ho_{n} (h, \cosh \tilde{\chi})
\right. \nonumber  \\  & & \left.  \hspace{1cm}
+\theta(\tilde{\chi}'-\tilde{\chi}) Jo_{n} (h,\cosh \tilde{\chi})\,
Ho_{n} (h,\cosh \tilde{\chi}')  \right]  \left. \frac{}{} \right\}
\label{21}
\end{eqnarray}
where $ h= ka $ ,
$$ Me_{n}= \int_0^{2\pi} |Se_{n}|^2
d\Theta, \hspace{5mm} Mo_{n} =\int_0^{2\pi} |So_{n}|^2 d\Theta $$
are the normalization constants. The Heavyside unit step function is
denoted by $\theta$ and  $\tilde{Z}$ is the distance between two points.
Returning back to the original set of independent variables we note that
$ Z^2( \tilde{\chi},\Theta ) \equiv \tilde{Z}^2 = - Z^2(\chi, \Theta) $ and,
accordingly, in eq.(\ref{21}) the function $H_0$ goes over into $K_0$, the
modified Hankel function. Then the expression (\ref{green1}) takes the form
\begin{equation}
\label{22}
 G(x,x') =\frac{1}{(2\pi)^3} \int_0^{2\pi} d\phi \int^{\infty}_0 k \,
dk\, e^{ik Y \cos \phi} \, K_0 (kZ)
\end{equation}
and the distance functions $Z$ and $Y$ are given by
\begin{eqnarray}
 Z^2 & = &  r^2+a^2 \cos^2 (\theta-\phi)+r'^2+a^2 \cos^2 (\theta'- \phi)
 \nonumber\\ & &
-2 \left[ (r^2+a^2)^{1/2} (r'^2+a^2 )^{1/2}
\cos(\theta - \phi) \cos(\theta'- \phi) \right. \nonumber \\
& & \left. +  r r' \sin(\theta - \phi) \sin(\theta'- \phi) \right]
\label{23} \\ [2mm]
Y^2 & = & (y-y')^2+(z-z')^2.  \nonumber
\end{eqnarray}
We shall first perform the $k$ integration using the standard
integral \cite{gr}
\begin{eqnarray}
 \int_0^{\infty} k \, dk e^{ik \alpha} K_0(kZ) & = &
- \frac{\alpha}{(\alpha^2+ Z^2)^{3/2} } \left[ \ln \left(\frac{\alpha}{Z} +
 \sqrt{ \frac{\alpha^2}{Z^2} +1 } \right) + i \frac{\pi}{2} \right]
 \nonumber \\  & &  + \,\,
  \frac{1}{\alpha^2+Z^2}
\label{25}
\end{eqnarray}
with  $ \alpha = Z \cos \phi $. Substituting this expression into
eq.(\ref{22}) and carrying out the definite integral
over an angular variable $\phi $ we find the final closed expression for
the Green function
\begin{equation}
\label{30}  G(x,x') = \frac{1}{64 \pi^2} \frac{F}{F^2+Y^2 E},
\end{equation}
with
\begin{eqnarray}
F^2 & = & 2  \left\{
 \left( \cos(\theta'-\theta) -1 \right)^2 [ X^4+a^2 X^2+ X^2 g] \right.
   + 6 x^2 X^2   \nonumber \\
 && \left.  + \left( \cos(\theta'-\theta) + 1 \right)^2
 [x^4+a^2 x^2-x^2 g]  - 2 x^2 X^2 \cos^2(\theta'-\theta)
          \right\} , \nonumber \\ [2mm]
E & = &  \cos(\theta'-\theta) (-X^2+x^2 -g) +2X^2+2x^2+a^2  \nonumber \\
      &&   -[a^2 \cos(\theta'-\theta) -g +X^2-x^2]
       \cos(\theta+\theta')
\label {32}
\end{eqnarray}
where we have used the notations
$$ X  =  \frac{r+r'}{2}\,, \;\;\; x=\frac{r'-r}{2}\,,
\;\;\; g  =  [(a^2+X^2)^2+2x^2(a^2-X^2)+x^4]^{1/2}. $$
This Green function can be used to
evaluate the vacuum expectation value of the stress-energy
tensor for a massless scalar field. Using the point-splitting
procedure  we find a finite expression for renormalized Green function
by substraction from (\ref{30}) its singular part
\cite{DW}. This gives us the vacuum expectation value of
the stress-energy tensor
\begin{equation}
< T_{\mu}^{\;\nu}(x) > = \frac{1}{1536 \pi^2}
\frac{a^2}{16\, r^2}\, \frac{28 r^2- 5 a^2}{(r^2+ a^2 )^{3}}
\left( \begin{array}{cccc} 3 & 0 & 0 & 0\\
                    0 & -1 & 0 & 0 \\
                    0 & 0 & -1 & 0 \\
                    0 & 0 & 0 & -1\end{array}  \right)
\label{333}
\end{equation}
at the coincidence points.

\section{Conclusion}

     The gravitational instanton metrics generated by classical minimal
surfaces, the catenoid and the helicoid, are Euclidean Bianchi
Type $VI_0$ and $VII_0$ metrics respectively.
In this paper we have discussed some properties
of the gravitational instanton that corresponds to the helicoid. We
have shown that it is an anti-self-dual solution of Petrov Type $I$
and admits a Killing tensor which leads to the solution of the
Hamilton-Jacobi and Laplace equations by separation of variables.
The interpretation of the self-dual metric generated by the helicoid
as a gravitational instanton is problematic due to its incompleteness
and the curvature singularity at $ r=0$. The final resolution of
this issue must await an analysis of the global structure of this
exact solution. However, hyper-K\"ahler instanton metrics sometimes admit
an $M$-theory resolution \cite{ket2} and are therefore of interest in
supergravity even though they are singular and not complete.
The remarkable symmetry properties of this metric enables us to obtain the
scalar Green's function by separation of variables and calculate
the vacuum expectation value of the stress-energy tensor.

\section{Acknowledgments}

  We thank G. W. Gibbons for many interesting and helpful conversations.
We thank also U. Camc{\i} for his interest in this work.

\pagebreak

\begin{center}
Figure Caption\\
Figure 1.  Helicoid in $S^2 \times S^1$.
\end{center}
\noindent
The unit disk at the cross-section represents $S^2$ and its geodesics
consist of great circles that intersect the boundary at right angles.
The helicoid surface is generated by one such geodesic, that is the surface
is {\it ruled} by this geodesic of $S^2$.


\begin{thebibliography}{99}

\bibitem{bpst} Belavin A A, Polyakov A M, Schwarz A S and
   Tyupkin Yu S 1975 Phys. Lett. {\bf 59B}  85

\bibitem{th}  t'Hooft G 1976 Phys. Rev. Lett. {\bf 37} 8

\bibitem{jnr}  Jackiw R, Nohl C and Rebbi C  1977 Phys. Rev. {\bf D15}
1642

\bibitem{ket2} Ketov S V {\it Analytic tools to brane technology
in $ N=2$ gauge theories with matter}, hep-th/9806009.

\bibitem{gib} Gibbons G W {\it Gravitation and Relativity: At the turn of
the Millennium} Proceedings of the GR-15 Conference, eds. Dadhich N and
Narlikar J. 1997 India

\bibitem{sw} Seiberg N and Witten E 1994 Nucl. Phys. {\bf B 426} 19,
{\it ibid} {\bf B 431} 484

\bibitem{cihan} Saclioglu C 1998 {\it Monopole Condensation in
Seiberg-Witten $4$-Manifold Theory} FGI preprint.

\bibitem{h}  Hawking S W 1977 Phys. Lett. {\bf 60A} 81

\bibitem{gh}  Gibbons G W and Hawking S W 1978 Phys. Lett. B {\bf 78} 430

\bibitem{gp}  Gibbons G W and Pope C N 1978 Commun. Math. Phys. {\bf 61}
239

\bibitem{yau}  Yau  S T  1978 Comm. Pure and Appl. Math. {\bf 31} 339

\bibitem{ahs}  Atiyah M F, Hitchin N and Singer I M 1978 Proc. Roy. Soc. A
{\bf 362} 425

\bibitem{gh1} Gibbons G W and Hawking S W 1977 Phys. Rev. {\bf D15} 2752

\bibitem{eh}  Eguchi T and  Hanson A J 1978 Phys. Lett. {\bf 74B} 249

\bibitem{egh}  Eguchi T, Gilkey P H and  Hanson A J 1980 Phys. Rep. C
{\bf 66} 213

\bibitem{min}   Nutku Y 1996 Phys. Rev. Lett. {\bf 77} 4702

\bibitem{comtet}  Comtet A 1978 Phys. Rev. D {\bf 18} 3890

\bibitem{jorgens}  J\"orgens K  1954  Math. Annalen {\bf 127} 130

\bibitem{gsvy}  Greene B R,  Shapere A,  Vafa C and  Yau S T 1990
Nuc. Phys. B {\bf 337} 1

\bibitem{gorr}  Gibbons G W, Ortiz M E and Ruiz Ruiz F 1990 Phys. Lett. B
{\bf 240} 50

\bibitem{weier}  Weierstrass K 1903 {\it Math. Werke} Vol. 3

\bibitem{akn}  Aliev A N, Kalayc{\i} J and Nutku Y 1997 Phys. Rev D
{\bf 56} 1332

\bibitem{previous}  Aliev A N and  Nutku Y  1999 Class. Quant. Grav.
{\bf 16}; gr-qc/9805006

\bibitem{mac} MacCallum M A H 1979 in {\it General Relativity,
Einstein Centenary Survey} ed Hawking S W and Israel W (Cambridge,
Cambridge University Press)

\bibitem{bgpp} Belinski V A, Gibbons G W, Page D N and Pope C N 1978
Phys. Lett. B {\bf 76} 433

\bibitem{iw} Inonu E and Wigner E P 1953 Proc. Nat. Acad. Sci. {\bf 39} 510

\bibitem{gru}  Gibbons G W and Ruback P J  1988 Commun. Math. Phys.{\bf 115}
267

\bibitem{wp}  Walker M and Penrose R 1970 Commun. Math. Phys. {\bf 18} 265

\bibitem{car1}  Carter B  1968  Phys. Rev. {\bf 10} 1559

\bibitem{mf}  Morse P M and Feshbach H  1953  {\it Methods of Theoretical
Physics} ( New York,  McGraw-Hill)

\bibitem{gr}  Gradshtein I S and Ryzhik I M 1965 {\it Table of Integrals,
Series and Products} (New York, Academic Press )

\bibitem{DW}  DeWitt B S 1975 Phys. Reports {\bf 19C}  297


\end{thebibliography}
\end{document}